\documentclass{elsart}
\usepackage{graphicx}

\usepackage{amssymb}

\begin{document}

\begin{frontmatter}

\title{Combined effects of disorders and electron-electron interactions
upon metal-insulator transition in 2D non-bipartite lattice}

\author{Jiaxiang Wang, Sabre Kais}

\address{Department of Chemistry, Purdue University West Lafayette, IN 47907}

\begin{abstract}
We examined the characteristics of a metal-insulator transition on a two-dimensional
non-bipartite lattice when both disorders and electron interactions are present. 
Using a real-space renormalization group method and finite-size scaling analysis
we have found that after the  disordered system transits from the insulating state
to the metal one, it can reenter an insulating state if there is a further increase in the 
electron-electron interactions.
\end{abstract}

\end{frontmatter}

Ever since the discovery of the metallic properties on effectively 2D electron systems in
metal-oxide-semiconductor field-effect
transistor (MOSEF's) \cite{krav}, there have appeared a spate of discussions and investigations over its
underlying physical mechanism \cite{abra}. In experiments, as the carrier density $n$ is increased
above a critical density $n_c$, 
the conductivity $\sigma _{dc}$ decreases upon lowering the temperature $T$ (typical of an insulator)
and as $n<n_c$, $\sigma _{dc}$ increases upon lowering $T$ (typical of a conductor). Because the data can be
scaled onto two curves, one for the metal and the other for the insulator,this phenomenon is seen as evidence for
the occurrence of a quantum-phase transition, namely the metal-insulator transition (MIT), with carrier density as the
tuning parameter. But there are     also other viewpoints, supported by some  
experiments \cite{papa,yaish}, maintaining that there is no 2D MIT at zero temperature and the metallic
behavior is attributed to the conventional, though nontrivial, electron transport. For example, two of these models
are based upon electron scattering by impurities
\cite{alts} and the effect of temperature dependent screening \cite{das,dewalle}. Thus there is, as
yet, no general consensus about the origin
of the metallic behavior and it remains a controversial topic.   
  
According to the conventional one-parameter scaling theory of noninteracting electrons \cite{abra1},
any amount of disorders will localize the 2-D electronic state and make it insulating. Moreover, the metallic state
has been seen in highly mobility structures at low carrier concentrations. Thus it is expected that the combining
effect of both electron-electron interactions and the disorders are very likely to play an important role. In fact,
the role of electron-electron interactions in disordered systems was recognized long ago by Finkelstein \cite{fink}
and Castellanic \cite{cast}, who treated the disorder in lowest order, where all interactions contributed
t$\frac {}{}$o the leading logarithmic behavior were summed. Recently, Si and Varma \cite{si} calculated a correction
to the compressibility of a disordered system by considering the ring diagrams. Because of the perturbation
nature of these methods, they can not deal with the regime with the disorder and the interactions having comparable
magnitudes. Hence most recent work has used exact diagonalization \cite{kot,raphy} and Monte Carlo methods
\cite{den} in the investigations. As we know, both methods suffer from intensive calculations and it
is very difficult to apply them to large-size systems. To resolve this difficulty, we resort to the
real-space block renormalization    
group (BRG) method\cite{ma}. Although this method has uncontrollable approximations, it can give us
many qualitative and insightful results and show us the direction for further more accurate work.        
The model we use is the Anderson-Hubbard model, which is the most simple one to include the essential
ingredients for our purpose, i.e. the disorder and the electron-electron interactions. We will use this model to
study the charge gap and the electron localization and delocalization on a triangular half-filled lattice as
shown in Fig.1. Normally there are two kinds of disorders. One is the site disorder related to the site fluctuations
and the other is the bond disorder related to the hopping terms. Here we will only consider the former one. The
Hamiltonian for Anderson-Hubbard model in our case can be written as,      
  
\begin{equation}   
H=-t\sum_{<i,j>,\sigma }[c_{i\sigma    
}^{+}c_{j\sigma    
}+H.c.]+U\sum_i(n_{i\uparrow }-\frac   
12)(n_{i\downarrow }-\frac   
12)-\sum_i\varepsilon _i(n_{i\uparrow  
}+n_{i\downarrow   
}),\label{1}
\end{equation}     
where $t$ is the nearest-neighbor hopping (exchange coupling) term, $U$ is the local repulsive
interaction and $\mu $ is the chemical potential. $c_{i\sigma }^{+}(c_{i\sigma })$ creates (annihilates) an electron with
spin $ \sigma $ in the valence orbital of the dot
located at site $i$; the corresponding number operator is  $n_{i\sigma}=c_{i\sigma}^{+}c_{i\sigma}$.
$<\cdot\cdot\cdot>$ on the first sum in Eq. (1) indicates that summation is restricted to nearest-neighbor dots. H.c. denotes
the Hermitian conjugate. Note that this model Hamiltonian allows only one orbital per dot. That orbital can be empty
or accommodate one or two electrons. $U$ is the repulsion of two electrons (of opposite spins) placed in
the same dot. $\varepsilon _i$ measures the fluctuation of the site energies. It is assumed that $\varepsilon _i$
follows a Gaussian distribution with the width to be $W$, i.e., 
  
\begin{equation}   
P(\varepsilon _i)=\frac 1{\sqrt{2\pi   
W}}e^{-\frac{(\varepsilon _i-\overline{\varepsilon })^2}{2W}},  \label{2}     
\end{equation}     
in which the bar over $\varepsilon $ means its average value. Initially, we use $\overline{\varepsilon
}=0.$   
  
The essence of the BRG method is to map the above many-particle Hamiltonian on a lattice to a new one
with fewer degrees of freedom and with the same low-lying energy levels \cite{book}. Then the mapping is repeated
leading to a final Hamiltonian of a seven-site hexagonal array for which we obtain an exact numerical solution.
  
When there is no disorder, namely $\varepsilon _i=0$, the procedure can be summarized into three steps:
First divide the $N$--site lattice into appropriate $n_s$--site blocks labeled by $p$ ($p$=1,2,...,
$N$/$n_s$) and separate the Hamiltonian $H$ into a intrablock part $H_B$ and an interblock $H_{IB}$, 
\begin{equation}   
H=H_B+H_{IB}={\sum\limits_p}H_p+{\sum\limits_{{\left\langle{p,{p}^{\prime }}\right\rangle  
}}}V_{p,{p}^{\prime }}, \label{3}    
\end{equation}     
\noindent
where $H_p$ is the Hamiltonian (1) for a given   
block and the interblock $%
p, p^{\prime }$ coupling is defined in Eq. (4)   
below.   
  
The second step is to solve $H_p$ exactly for the eigenvalues $E_{p_i}$ and eigenfunctions $\Phi _{pi}
(i=1,2,...,4^{n_s}).$ 
Then the eigenfunctions of $H_B$ are constructed by direct multiplication of $\Phi _{pi}$ \textsf{.}
The last step is to treat each block as one site on a new lattice and the correlations between blocks as hopping
interactions.      
  
The original Hilbert space has four states per site. If we are only concerned with lower lying states
of the system as when studying the metal-insulator-transition \cite{imada}, it is not necessary to keep all the
states for a block.    
  
To make the new Hamiltonian tractable, the reduction in size should not be accompanied by a
proliferation of new couplings. Then one
can use an iteration procedure to solve the model. To achieve this, it is necessary to keep only 4
states in step 2. Their energies     
are $E_i$ ($i$=1,2,3,4). In order to avoid proliferation of additional couplings in the new
Hamiltonian, the four states kept from the block cannot be arbitrarily chosen. Some definite conditions as discussed in Ref.
\cite{jwang} must be satisfied. For example, the states must belong to the same irreducible representation of
${C_6\nu }$ symmetry group of the lattice. In particular, in order to copy the intrasite structure of the old
Hamiltonian, $E_3=E_4$ is a necessary condition. Furthermore, particle-hole symmetry of a half-filled lattice
requires that $E_1=$ $E_2$. Further restrictions follow from the need to make extra couplings vanish. Operators in
the  
truncated basis are denoted by a prime so that the interblock coupling of Eq. (2) is

\begin{equation}   
V_{pp^{\prime }}=\nu \lambda ^2t\sum_\sigma      
c_{p\sigma }^{\prime  
+}c_{p^{\prime }\sigma }^{\prime },  \label{4}   
\end{equation}     
where $\nu $ represents the number of couplings between neighboring blocks. The coupling strength for
the border sites of a block by {$\lambda $} and the renormalization group equation for the coupling strength is 
\begin{equation}   
t^{\prime }=\nu \lambda ^2t,  \label{5}
\end{equation}     
The other renormalization relation is  
\begin{equation}   
U^{\prime }=2(E_1-E_2).  \label{6}      
\end{equation}     
  
Once we introduce the disorders, because there is no exact particle-hole symmetry any more, the
parameters $t,$ $U$ and $\varepsilon _i$ have to be renormalized on average. In details, let us use $\alpha $ and $\beta $
to be the block indices. Then for one block $\alpha ,$ we can have 
  
\begin{eqnarray}   
U^\alpha  &=&E_1^\alpha +E_2^\alpha -2E_3^\alpha 
,  \label{7} \\    
\varepsilon ^\alpha  &=&E_2^\alpha -E_1^\alpha . 
\label{8}
\end{eqnarray}     
After the renormalization, the new energies $\varepsilon ^\alpha $ do not obey a Gaussian distribution.
In order to iterate the RG, as in Ref.\cite{ma}, we adopt the following procedures: 
  
(1) $\varepsilon ^\alpha $ is forced back into a Gaussian distribution with the new width 
  
\begin{equation}   
W^{\prime }=\overline{(\varepsilon ^\alpha
)^2}-(\overline{\varepsilon ^\alpha })^2. 
\label{9}
\end{equation}     
The new Gaussian is not centered at zero since there will be a constant shift due to the electron
interactions. But we can still take it to be zero by formally introducing the chemical potential.
  
(2) $U^\alpha $ is forced back to be constant.   
  
\begin{equation}   
U^{\prime }=\overline{U^\alpha }.  \label{10}    
\end{equation}     
  
(3) To get the renormalized hopping term, we will have to consider all the possible non-zero average
values of the coupling between the block states. For two neighboring blocks, there are 4 possibilities.     
  
\begin{eqnarray}   
t_1^{\alpha \beta } &=&t<\uparrow ^\alpha 0^\beta
|\sum_{<\alpha     
i,\beta j>,\sigma }[c_{\alpha i\sigma  
i,\beta j>}^{+}c_{\beta j\sigma 
}]|\uparrow ^\beta 
0^\alpha >,  \label{11} \\   
t_2^{\alpha \beta } &=&t<\uparrow \downarrow     
^\alpha 0^\beta    
|\sum_{<\alpha i,\beta j>,\sigma }[c_{\alpha     
i\sigma }^{+}c_{\beta 
j\sigma }]|\uparrow
^\beta \downarrow ^\alpha >,  \label{12} \\      
t_3^{\alpha \beta } &=&t<\uparrow ^\alpha 
\downarrow ^\beta  
|\sum_{<\alpha i,\beta j>,\sigma }[c_{\alpha     
i\sigma }^{+}c_{\beta 
j\sigma }]|\uparrow
\downarrow ^\beta 0^\alpha >,  \label{13} \\     
t_4^{\alpha \beta } &=&t<\uparrow \downarrow     
^\alpha \downarrow 
^\beta |\sum_{<\alpha i,\beta j>,\sigma
}[c_{\alpha i\sigma
}^{+}c_{\beta j\sigma }]|\uparrow \downarrow     
^\beta \downarrow  
^\alpha >.  \label{14}
\end{eqnarray}     
We also force the distribution of $t^{\alpha \beta }$ into a Gaussian with mean 
  
\begin{equation}   
t^{\prime }=\overline{t_i^{\alpha \beta }},      
\label{15}  
\end{equation}     
and width
\begin{equation}   
t_2^{\prime }=\overline{\left( t_i^{\alpha \beta 
}\right) 
^2}-\left( \overline{t_i^{\alpha \beta }}\right) 
^2.  \label{16}    
\end{equation}     
  
In principle, to use more blocks for averaging is always desirable. In our work, we use 7 unit blocks,
which themselves form a hexagonal block. 
  
For clean systems, we have shown in our former work \cite{jwang} the existence of Mott MIT at $U=12.5$
by investigating the charge gap $\triangle _p$, which can be expresses as the limiting value of the
renormalized $U$,     
  
\begin{equation}   
\Delta _g={lim}_{n\rightarrow \infty   
}\;\;U^{\prime (n)}.  
\label{17}  
\end{equation}     
For disordered systems, as we know from the orthodox scaling theory \cite{abra1}, the noninteracting
electrons will be localized, which leads to another kind of insulator, namely the Anderson insulator. The localization
feature can be exposed by the  inverse participation rate (IPR) \cite{weg} $\xi$ defined as
  
\begin{equation}   
\xi ^{(4)}=\frac{N_e}{{\sum\limits_i}\left|      
\left\langle \psi  
\left| \widehat{n}_i\right| \psi \right\rangle   
\right| ^4},
\label{18}  
\end{equation}     
where $N_e$ is the total number of the electrons, which is equal to the site number $N_s$ and $\psi $
the system wave function. For totally localized electronic states in the half-filled system, half of the sites
have $\left\langle \psi \left| \widehat{n}_i\right|\psi \right\rangle=2.$ On the other hand, when the
electron is totally delocalized, the average number of electrons per site should be $\left\langle \psi \left|
\widehat{n}_i\right| \psi \right\rangle =1.$ Hence the IPR $\xi ^{(4)}$ will satisfy $0.125\leq \xi^{(4)}\leq 1$ between the
above two limiting cases.
  
In Fig.2, we demonstrated how the disorders influence the Mott MIT. It is interesting to note that the
disorder shows the effect of destablizing the Mott insulating state, namely, as we increase the disorder, part
of the insulating region appears ''melted'' 
into a ''metallic'' state, in which the charge gap is nearly zero. At first sight, this feature quite
contradicts the generally accepted viewpoints that the disorder should stabilize the insulating state. But actually
they are talking about different interaction regimes with two different MIT mechanism. Our result is obtained in
a strong coupling regime with the charge gap to measure MIT (Mott MIT). As will be shown below, for weak
electron-electron interactions, the disorders really stabilize the insulating state with $\xi ^{(4)}$ to signature
MIT(Anderson MIT). 
  
Fig.3 displays the evolving of electron localization and delocalization as we tune both the disorder
and electron-electron interaction in the whole coupling regime. It is easy to see that all the curves
follow almost the same variation patterns as we increase $U$ for fixed disorder $W$. Generally speaking, there are three
stages.      
  
1) When $U$ is very small, the electron-electron interactions will help with the electron
delocalization. This is physically understandable since the on-site repulsion will  tend to rearrange the charges and make the
charge-density more homogeneous. This result is consistent with the findings by Ma \cite{ma} and
Caldara \cite{caldara}, but is different from the results from Berkovits \cite{berkovits}and Benenti
\cite{benenti}. The reason for this difference might be because Berkovits and Benenti use spinless electrons and only
consider the neighbor-site repulsion, which does not have the homogenizing effect like the on-site interaction.  
  
2) As we increase $U$ further to about $U\sim W$, the electron become maximally delocalized and becomes
localized again when $U>W$. The delocalization peak in the diagram is very apparent when $W$ is,
again when $U$ is, for example, bigger than 10. If we interpret the shallow region around the peak to be in a metallic state, the system will then
experience here a three-phase transition: insulator-metal-insulator, which is quite consistent with the experimental
findings \cite{and,ham}. Another interesting fact we might note is that, as the disorders increases, the
metallic region moves further to the right of the axis, which implies that the initial insulating state becomes more
stabilized. This is what we have mentioned before.    
  
3) After the two phase transitions,  the system continues to exhibit Anderson insulating behavior until
finally it evolves into a Mott insulating state with very strong electron couplings. In this regime, every site
has one electron on average because of the strong on-site electron interactions.  
  
From the above discussion, we should be able to summarize four different electronic features as $U$
increases from zero to infinity, namely, Anderson Insulator-Metal-Anderson Insulator-Mott Insulator. This full
scenario from weak to strong coupling is quite consistent with the present experiments \cite{and,ham}.   
  
One of the advantages of the BRG method is that it can be adapted easily to carry out finite-size
scaling analysis of the system \cite{jwang1}. Thus more fundamental physics can be explored and displayed.  
  
Fig.4 presents our calculation results with respect to the charge gap for each fixed disorder. It is
very interesting to note that as we increase the disorders, the finite-size scaling analysis provide  us one
critical point while $W$ is small, which clearly represents quantum phase transition. Then the system enters
some disorder/interaction range where no clear critical transition
point can be identified. We call this region a  ''mixed'' region since the two pure phases that is
necessary to characterize the phase transition are difficult to define. this is demonstrated by the absence of a
clear crossing point from the finite-size scaling analysis. As the disorders become stronger, two critical points are
evolved out of the mixed region. From here, we can easily identify the two phase transitions as discussed above,
i.e. insulator-metal and metal-insulator transitions. But the last phase transition, Anderson insulator-Mott
insulator is still absent in the finite-size scaling diagrams up to the magnitude of the disorders we have
considered. It should be expected that one more critical point might appear if stronger disorders are considered. More
extensive calculations are needed before a final conclusion can be drawn.      
  
In summary, by using the BRG technique, we have investigated in detail the delicate interplay between
disorders and electron-electron interactions on a half-filled triangular quantum dot lattice. The
insulator-metal-insulator transition has been well demonstrated by the finite-size scaling analysis. \\    
    
Acknowledgments \\
We would like to acknowledge the financial support of the Office of Naval Research (ONR).

\newpage

\begin{figure}[bt]
\includegraphics[clip, viewport=0 200 460 580, width=0.8\textwidth]{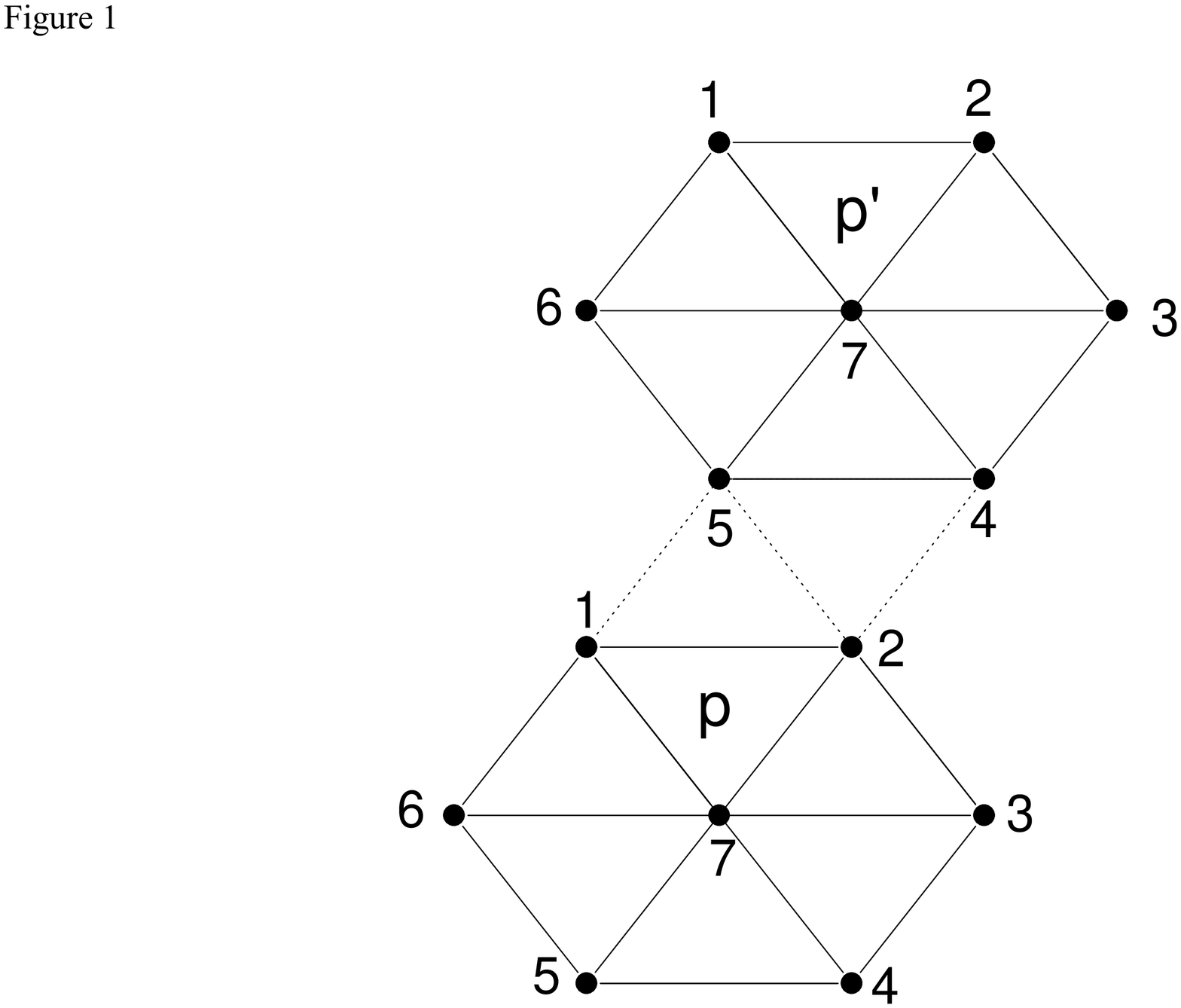} 
\caption{Schematic diagram of the triangular lattice with hexagonal blocks. Only two neighboring blocks
$p$    
and $p^{\prime}$ are drawn here. The dotted lines represent the interblock interactions and the solid
lines the intrablock couplings.}
\label{fig.1}
\end{figure}

\begin{figure}[bt]
\includegraphics[clip, viewport=60 370 460 610, width=0.8\textwidth]{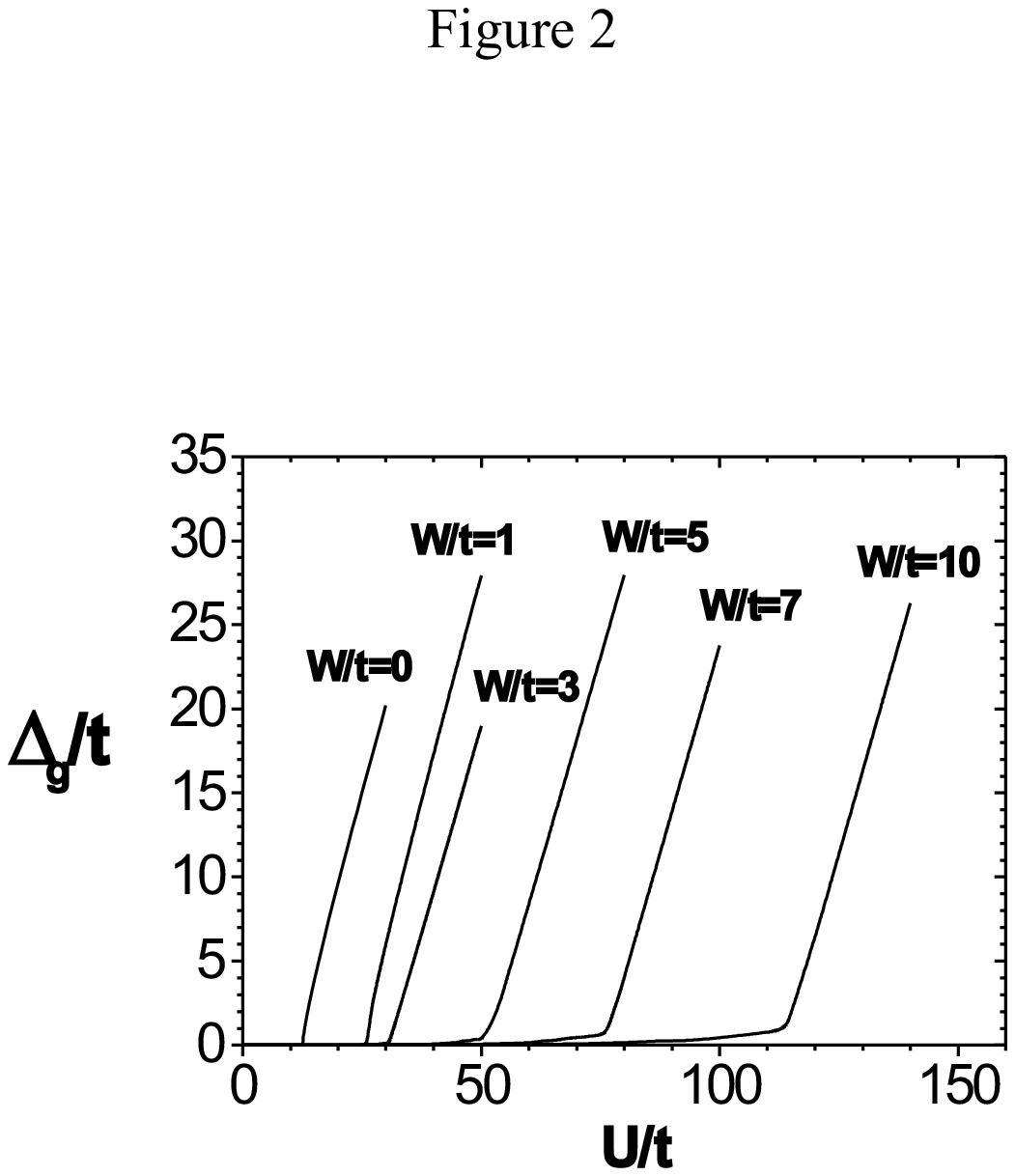} 
\caption{Influence of the disorder (measure by W/t) over the charge gap $ \triangle _g/$t dependence
upon the  
electron-electron interactions (U/t).}
\label{fig.2}
\end{figure}

\begin{figure}[bt]
\includegraphics[clip, viewport=0 160 460 700, width=\textwidth]{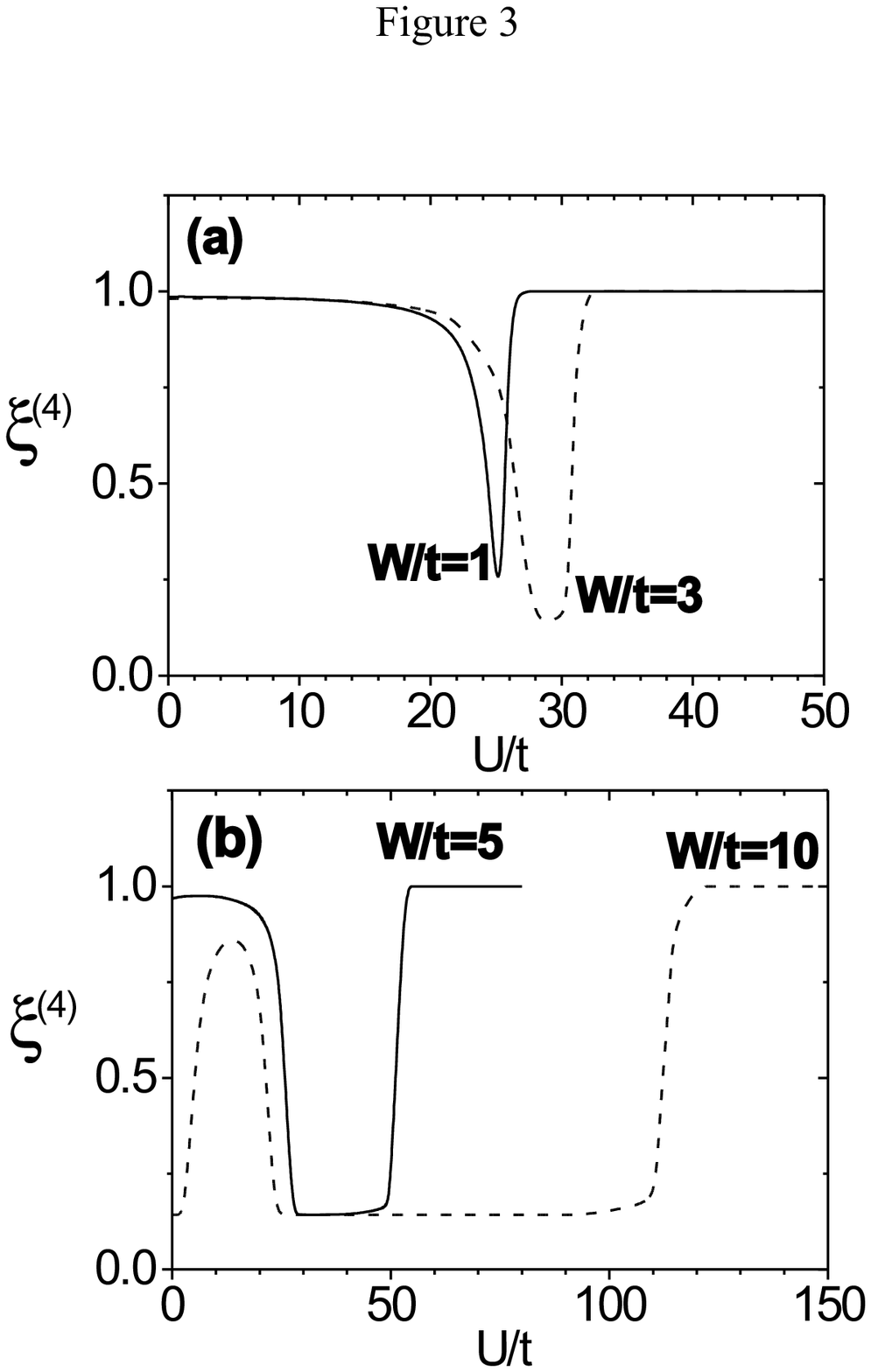} 
\caption{Variation of the inverse participation rate (IPR, $\xi^{(4)})$ against the electron-electron  
interactions for different disorders with (a) W/t=1(solid line), 3(dashed line) and (b) W/t=5(solid
line), 10(dashed line). }
\label{fig.3}
\end{figure}

\begin{figure}[bt]
\includegraphics[clip, viewport=0 100 460 680, width=\textwidth]{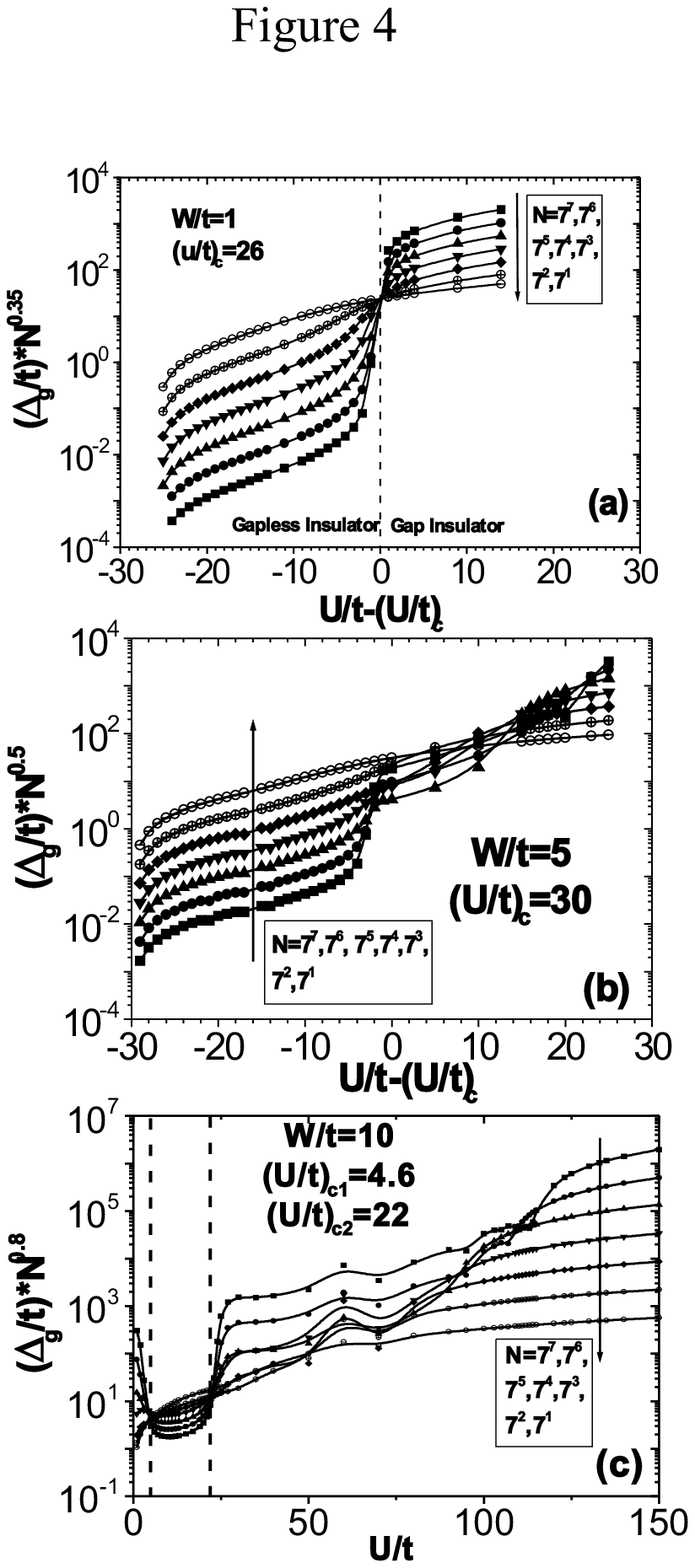} 
\caption{Finite-size scaling analysis over the charge gap $\triangle _g$/t against electron-electron   
interactions for fixed disorders. (a) W/t=1. (b) W/t=5. (c) W/t=10.}
\label{fig.4}
\end{figure}


\begin{thebibliography}{00}
\bibitem{krav} S. V. Kravchenko, G. V. Kravchenko, J. E. Furneaux, V. M. Pudalov, M. D'Iorio, 
Phys. Rev. B 50 (1994) 8039.
     
\bibitem{abra} E. Abrahams, S. V. Kravchenko, M. P. Sarachik, Rev. Mod. Phys. 73 (2001) 251 and
references therein.   
     
\bibitem{papa} S. J. Papadakis S. J., E. P. de Poortere, H. C. Manoharan, M. Shayegan, E. Winkler,
 Science 283 (1999) 2056.
   
\bibitem{yaish} Y. Yaish, O. Prus, E. Buchstab, S. Shapira, Y. G. Ben, U. Sivan, A. Stern,
  Phys. Rev. Lett. 84 (2000) 4954.
     
\bibitem{alts} B. L. Altshuler, D. L. Maslov, Phys. Rev. Lett. 82 (1999) 145.     
     
\bibitem{das} S. Das Sarma, E. H. Hwang, Phys. Rev. Lett. 83 (1999) 164.

\bibitem{dewalle} A. L. Dewalle, M. Pepper, C. J. B. Ford, E. H. Hwang, S. D. Sarma, D. J. Paul, G.
Redmond, 
Phys. Rev.B 66 (2002) 075324.      
     
\bibitem{abra1} E. Abrahams, P. W. Anderson, D. C. Licciardello, T. V. Ramakrishnan,Phys. Rev. Lett. 42
(1979) 673.     
     
\bibitem{fink} A. M. Finkelstein, Z. Phys. B: Condens. Matter 56 (1984) 189.      
     
\bibitem{cast} C. Castellani, C. Di Castro, P. A. Lee, M. Ma, Phys. Rev. B 30 (1984) 527.
     
\bibitem{si} Q. Si, C. M. Varma, Phys. Rev. Lett. 81 (1998) 4951; Physica B 259 (1999) 708.    
     
\bibitem{kot} R. Kotlvar, S. D. Sarma, Phys. Rev. Lett. 86 (2001) 2388.  
     
\bibitem{raphy} F. Remacle, R. D. Levine, Chemphyschem 2 (2001) 20; PNAS 97 (2000) 553.
     
\bibitem{den} P. J. H. Denteneer, R. T. Scalettar, N. Trivedi, Phys. Rev. Lett. 83 (1999) 4610.      
     
\bibitem{jwang} J. X. Wang, S. Kais, R. D. Levine, Inter. J. Mol. Sci. 3 (2002) 4.  

\bibitem{weg} F. Wegner, Z. Physik B 36 (1980) 209.
     
\bibitem{imada} M. Imada, A. Fujimori, Y. Tokura, Rev. Mod. Phys. 70 (1998) 1039.
     
\bibitem{book} "Topics in current physics: Real-space renormalization". Editors: T. W. Burkhardt and J.
M. J. van Leeuwen.
Published by Springer-Verlag Berlin.
     
\bibitem{ma} M. Ma, Phys. Rev. B 26 (1982) 5097.    
     
\bibitem{caldara} G. Caldara, B. Srinivasan, D. L. Shepelyansky, Phys. Rev. B. 62 (2000) 10680.      
     
\bibitem{berkovits} R. Berkovits, J. W. Kantelhardt, Y. Avishai, S. Havlin, A. Bunde, Phys. Rev. B 63
(2001) 085102. 
     
\bibitem{benenti} G. Benenti, X. Waintal, J. Pichard, Phys. Rev. Lett. 83 (1999) 1826.
     
\bibitem{jwang1} J. X. Wang, S. Kais, Phys. Rev. B 66 (2002) 081101(R).    
     
\bibitem{and} A. Andresen, C. Prasad, F. Ge, L. -H. Lin, N. Aoki, K. Nakao, J. P. Bird, D. K. Ferry, Y.
Ochiai,
Phys. Rev. B 60 (1999) 16050.
     
\bibitem{ham} A. R. Hamilton, M. Y. Simmons, M. Pepper, E. H. Linfield, P. D. Rose, D. A. Ritchie,
Phys. Rev. Lett. 82 (1999) 1542.   


\end{thebibliography}
\end{document}